# A volatile polymer stamp for large-scale, etching-free, and ultraclean transfer and assembly of two-dimensional materials and its heterostructures

*Zhigao Dai[1], Yupeng Wang[1], Lu Liu[1], Junkai Deng[2, *], Wen-Xin Tang[3], Qingdong Ou[4], Ziyu Wang[4, 5], Md Hemayet Uddin[6], Guangyuan Si[6], Qianhui Zhang[7, *], Wenhui Duan[7], Michael S. Fuhrer[4, 8], Changxi Zheng[9, 10]*,*

[1]Faculty of Materials Science and Chemistry, China University of Geosciences, Wuhan, Hubei 430074, P. R. China.

[2]State Key Laboratory for Mechanical Behavior of Materials, Xi'an Jiaotong University, Xi'an, 710049, China.

[3]Electron Microscope Center, Chongqing University, Chongqing, 400044, China.

[4]ARC Centre of Excellence in Future Low-Energy Electronics Technologies (FLEET), Monash University, Clayton, VIC 3800, Australia.

[5]Institute of Materials Research and Engineering, Agency for Science Technology and Research (A*STAR), Singapore 138634, Singapore.

[6]Melbourne Centre for Nanofabrication, Clayton, VIC 3800, Australia.

[7]Department of Civil Engineering, Monash University, Clayton, VIC 3800, Australia.

[8]School of Physics and Astronomy, Monash University Clayton, VIC 3800, Australia.

[9]Key Laboratory for Quantum Materials of Zhejiang Province, Department of Physics, School of Science, Westlake University, 18 Shilongshan Road, Hangzhou 310024, Zhejiang Province, China

[10]Institute of Natural Sciences, Westlake Institute for Advanced Study, 18 Shilongshan Road, Hangzhou 310024, Zhejiang Province, China

*To whom correspondence should be addressed. Emails: junkai.deng@mail.xjtu.edu.cn; sherry.zhang1@monash.edu; zhengchangxi@westlake.edu.cn

**Abstract**

The intact transfer and assembly of two-dimensional (2D) materials and their heterostructures are critical for their integration into advanced electronic and optical devices. Herein, we report a facile technique called volatile polymer stamping (VPS) to achieve efficient transfer of 2D materials and assembly of large-scale heterojunctions with clean interfaces. The central feature of the VPS technique is the use of volatile polyphthalaldehyde (PPA) together with hydrophobic polystyrene (PS). While PS enables the direct delamination of 2D materials from hydrophilic substrates owing to water intercalation, PPA can protect 2D materials from solution attack and maintain their integrity during PS removal. Thereafter, PPA can be completely removed by thermal annealing at 180 °C. The proposed VPS technique overcomes the limitations of currently used transfer techniques, such as chemical etching during the delamination stage, solution tearing during cleaning, and contamination from polymer residues.

1. Introduction

The ability to pick and place two-dimensional (2D) materials has enabled the creation of van der Waals (vdW) heterostructures[1-4] which have significant potential for numerous electronic and optoelectronic applications.[5-9] In contrast to conventional epitaxial approaches, the saturated bonds on vdW surfaces enables the preparation of atomically precise stacks of different 2D materials without considering lattice mismatch.[10, 11] However, the lack of dangling bonds on vdW surfaces makes it difficult to directly grow large-area heterostructures by epitaxy.[12] At present, vdW heterostructures are mostly fabricated by preparing large-area 2D materials individually on different substrates and then stacking them directly using transfer techniques. Numerous transfer techniques have been developed over the years;[1],[13-17] however, heterostructures assembled using the current transfer methods suffer from various issues, such as material damage caused by substrate etching solutions like potassium hydroxide, ruptures caused by polymer removal solvents/solutions at the cleaning stage, and contamination from polymer residues.[18, 19] Recently, the transfer and stacking of 2D materials with a stamping layer such as PDMS[20], nickel film[21] and thermal release tape[22] has been reported. But they often require complex processing such as controlled lifting speed or low adhesion between the 2D materials and their substrates, which hinders their applicability.

Herein, we report a novel transfer method called volatile polymer stamping (VPS) using polyphthalaldehyde (PPA) as an adhesive transfer layer. PPA is a unique self-unzipping polymer; it readily vaporizes at a low temperature (150 °C) which is safe for many 2D materials. We used optical, electrical, and scanning probe methods to thoroughly characterize transferred transition metal dichalcogenide (TMD) crystals, films, and vdW heterostructures. It is demonstrated that the transferred materials retained excellent material integrity, and the surface quality was comparable to that of the as-grown samples. Polystyrene (PS)/PPA composite can also be adopted as a hydrophobic stamp for efficient fabrication of vdW heterostructures by simple stacking. Furthermore, an intact layer of PPA acts as a strong supporting layer during the solvent soaking step in the transfer process, enabling the transfer of large-scale suspended 2D structures. We used the VPS technique to transfer a monolayer $WS_2$ film onto a transmission electron microscopy (TEM) grid with a span of 150 µm (suspended region), which is the largest ever reported for monolayer TMDs.

## 2. Results and Discussion

### 2.1 Etching-free delamination of 2D materials in VPS Strategy.

The first step of transfer is the delamination of the 2D materials from their substrate. It is believed that delamination between $WS_2$ and sapphire is caused by water intercalation at the interface owing to the hydrophilic nature of the sapphire surface.[23-26] To understand the detailed transfer mechanism, we performed a time series of experiments of water soaking together with density functional theory (DFT) calculations. **Figure 1**a-h shows the optical and AFM images of the $WS_2$ crystal taken before and after each step of water soaking. Before soaking, the increased height due to ambient water intercalation appears only at the edge of the $WS_2$ crystal.[24, 27] After soaking in water for 15 min, intercalated water layers are visible along the atomic steps of the sapphire substrate; some layers are marked with white arrowheads in Figure 1f. After soaking for 30 min (Figure 1c and g), the number of intercalated layer increase. After soaking for 45 min, the crystal is partly torn by nitrogen gas blow-drying owing to the weak adhesion between $WS_2$ and sapphire after water intercalation (Figures 1d and h). It should be noted that except for the torn regions, the edge profile of the crystal remains unchanged. This suggests that water does not etch the $WS_2$ crystals. As shown in **Movie S1**, the peel-off can be completed within several seconds after the water droplet contacts the PS/PPA polymer stamp. However, our water intercalation experiments (Figures 1f-h) suggest that only a small number of regions show a noticeable increase in height within such a short time of water intercalation. Thus, we deduce that, after contact with the water droplet, a small amount of water molecules diffuses quickly and distributes uniformly below the $WS_2$ crystals. This efficiently weakens the adhesion between $WS_2$ and sapphire without noticeable height variations. This is supported by DFT calculations shown in Figure 1i and j, which compare the adsorption energies of $WS_2$ monolayers on sapphire at different concentrations of water molecule. As shown in Figure 1j, the absorption energy of $WS_2$ decreases dramatically from 1.145 eV to 0.981 eV with only 7% of sapphire surface covered by intercalated water molecules with the height of $WS_2$ increases by 1.13 Å due to the amount of intercalated water. The further water intercalation only deceases the absorption energy slightly while the height of $WS_2$ remains increasing substantially. More detailed values of the DFT calculations are given in Table S1. The calculations well explain that the fast delamination of 2D materials on sapphire is due to intercalation of a small amount of water molecules uniformly at the interface.

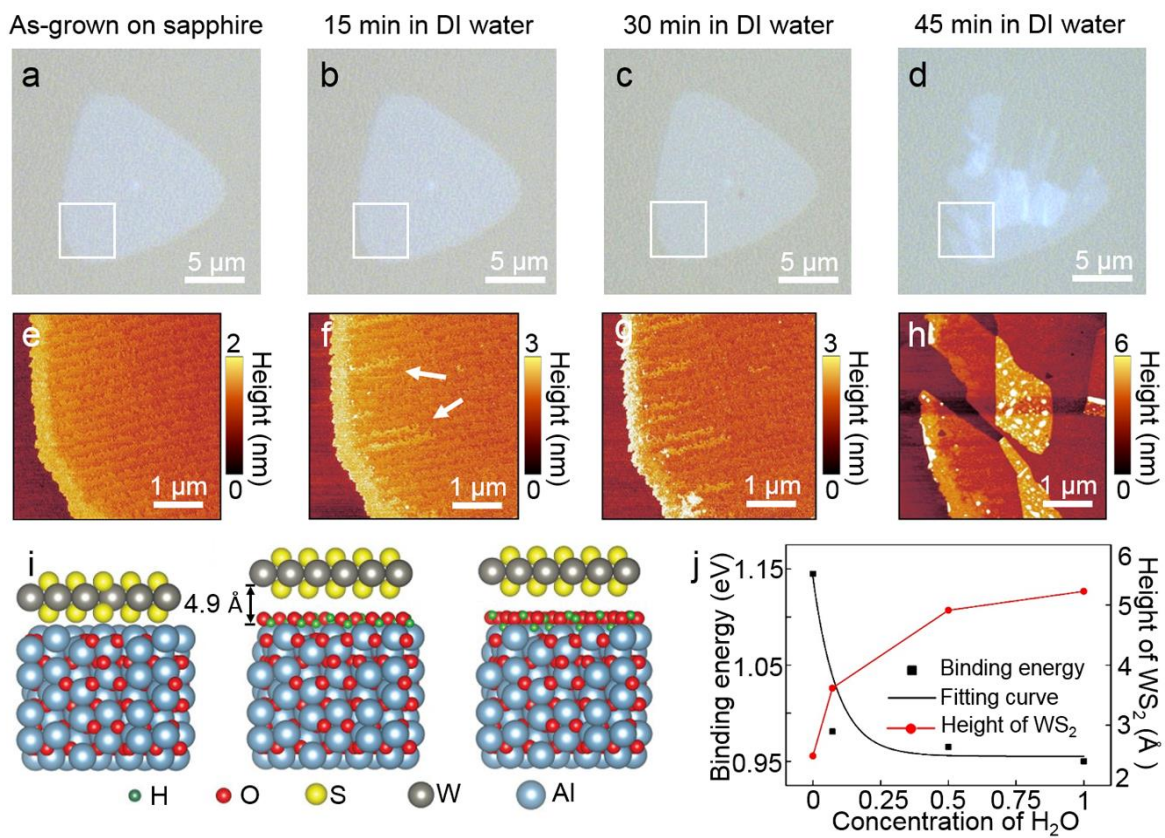

**Figure 1.** Water intercalation experiments and DFT calculations for the delamination process. (a-d) Optical micrographs of the CVD-grown $WS_2$ single crystal soaked in DI water for (a) 0 min, (b) 15 min, (c) 30 min, and (d) 45 min. (e-h) Topographic AFM images of the regions

marked with white squares in (a-d), respectively. (i) DFT calculations of monolayer $WS_2$ adsorption on sapphire surfaces with no water molecules, half covered by water molecules and fully covered with water molecules respectively. (j) The change of binding energy of $WS_2$ and height of $WS_2$ on the sapphire substrate with the increase of water molecules intercalated in between sapphire and $WS_2$ layers.

**2.2 Large-scale and ultraclean transfer of $WS_2$ crystals by VPS technique.**

The central feature of the proposed VPS technique is the use of PPA (**Figure 2**a), which is a volatile polymer evaporating at temperatures ≥ 150 °C. Such a low sublimation point enables thorough removal of the polymer residue with mild thermal annealing, thus minimizing heat damage to the transferred 2D materials. A relatively rigid PS layer was further spin-coated on top to strengthen the soft PPA layer. The reasons for selecting PS as the additional layer are as follows. (1) PS is a hydrophobic polymer; Gurarslan et al. has demonstrated the surface-energy-assisted direct transfer of chemical vapor deposition (CVD)-grown $MoS_2$ monolayer from sapphire substrate using only PS.[23] (2) The solubility of PPA in toluene is very low compared with that of PS, which enables the removal of the PS layer from PPA using toluene solution after transfer. (3) PPA has a higher adhesion than poly(methyl methacrylate) (PMMA) and PS (see Figure S1). The improved adhesion of PPA facilitates direct stacking of vdW monolayer materials by repeatedly using the volatile polymer stamp on a sapphire substrate with CVD-grown materials to form artificial heterostructures, as shown schematically in Figure 2a. The detailed mechanism of the direct stacking is discussed below.

The polymer removal method of the VPS technique is shown schematically in Figure 2b. The introduction of PPA in the VPS technique facilitates polymer removal by low-temperature (180 °C) thermal annealing rather than chemical solutions, thereby preventing rupture of the transferred materials by the chemical solution during the cleaning process. After stamping on the target substrate, the sample was first soaked in toluene solution for 30 min to completely dissolve the top PS layer, leaving an intact PPA layer on the deposited substrate to protect the transferred materials from the solution. Subsequently, the PPA layer was completely removed in a low-pressure tube furnace heated to 180 °C under flowing argon gas.

This two-step polymer removal method generates ultra-clean and well-preserved transferred material, as demonstrated by the $WS_2$ single crystal before and after transfer (Figure 2c-h). Figure 2c shows the photograph of the sapphire substrate with CVD-grown $WS_2$ crystals. The optical and atomic force microscopy (AFM) images of an as-grown WS$_2$ crystal are shown in Figures 2d and e, respectively. The details of the CVD process are provided in Methods and Ref. 22. It is evident from Figure 2e that the WS$_2$ surface is clean and atomically flat, following the atomic-height steps and atomically flat terraces of the underlying sapphire substrate with high precision. A slight increase in the height is observed around the crystal edge due to ambient water intercalation at the interface.[24, 27] Figure 2f shows the photograph of the SiO$_2$/Si substrate covered by an intact PPA layer after the removal of PS. The PPA layer contains the WS$_2$ crystals transferred from the sapphire substrate shown in Figure 2c. Figures 2g and h show the optical and AFM images of the transferred WS$_2$ crystal, respectively. The AFM image shows no evidence of polymer residue after the removal of PPA by thermal annealing. The transferred WS$_2$ features atomically flat morphology except a few wrinkles, which is induced by the wet transfer process. The height of wrinkles is within 10 nm, as indicated in the height profile of a representative wrinkle in Figure S2. This method can be adapted for wafer-scale material transfer (see Figure S3).

Furthermore, as the transfer process involves only deionized (DI) water and no etchant, the sapphire substrates after material transfer can be repeatedly used for CVD growth. Figures 2i-k show the photograph, optical image, and AFM image of CVD-grown WS$_2$ on a recycled sapphire substrate. The new crystals also exhibit the typical triangular faceting (Figure 2j), and the substrate surface retains the atomic steps and atomically flat terraces (Figure 2k). The distinct shape of the new crystals from the first-time grown crystals is likely due to the difference in surface energy of the sapphire substrate after high temperature annealing from the first CVD growth.[28, 29]

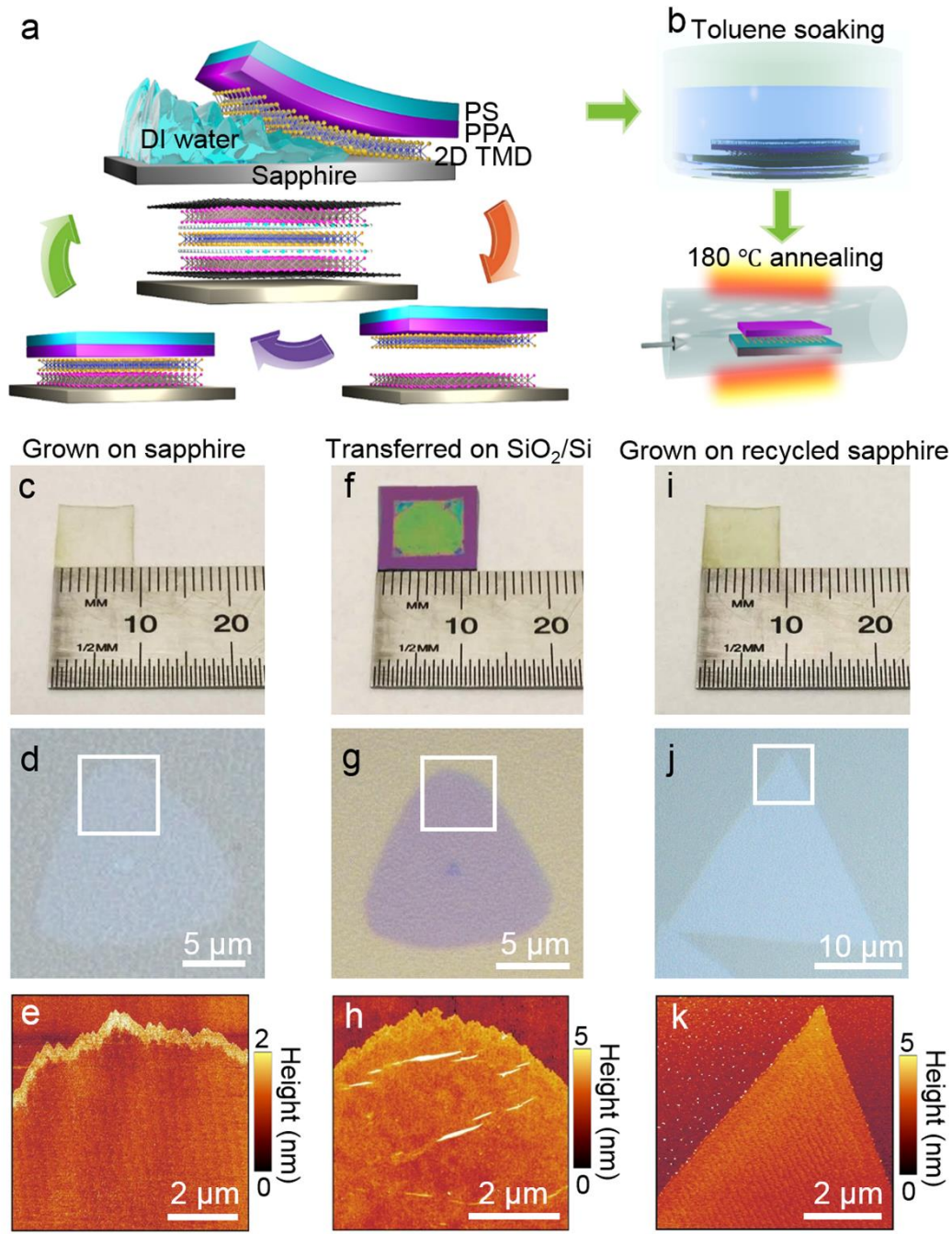

**Figure 2**. Transfer process using the VPS technique and morphology of $WS_2$ crystals. (a) Schematics of the VPS delamination and stamping process. (b) Schematics of the VPS polymer removal process. (c) Photograph of the CVD-grown $WS_2$ on pristine sapphire substrate. (d) Optical micrograph of the $WS_2$ single crystal in (c). (e) Topographic AFM image of the region marked with white square in (d). (f) Photograph of the $WS_2$ in (c) transferred on a $SiO_2$/Si substrate by VPS after the removal of PS with toluene. (g) Optical micrograph of the $WS_2$ single crystal in (f) after the removal of PPA. (h) Topographic AFM image of the region marked with white square in (g). The white features are wrinkles induced during the wet transfer

process. (i) Photograph of CVD-grown WS$_2$ on recycled sapphire substrate (from (c)). (j) Optical micrograph of the WS$_2$ single crystal in (i). (k) Topographic AFM image of the region marked with white square in (j).

We further illustrate the large-scale and ultraclean transfer of WS$_2$ crystals using VPS technique with optical, topological and electrical characterization. As shown in **Figure 3**a, after polymer removal, no polymer residues are observed in the WS$_2$ layer transferred on the SiO$_2$/Si substrate, and no imperfections such as tears, cracks, or folding are observed in the WS$_2$ crystals over a large area (560 × 150 μm$^2$). This indicates the multiple advantages of this transfer method over other strategies (see Figure S4 and Table S2). The high-magnification optical image (Figure 3b) shows that the transferred WS$_2$ layer is clean, and the Raman (Figure 3c) and photoluminescence (PL) (Figure 3d) maps indicate its uniform optical properties. The Raman and PL spectra of the WS$_2$ crystals transferred on SiO$_2$/Si are shown in Figure S5. We also compared the PL intensity of a WS$_2$ single crystal before and after the transfer, and the results are shown in Figure S6. The strong PL signal (approximately 7 times stronger) of WS$_2$ after transfer indicates that this transfer method efficiently preserves the optical properties.

Scanning probe microscopy was used to analyze the quality of the transferred WS$_2$ crystals deposited on an ultra-flat and clean sapphire substrate (Figure 3e) and a Pt-coated Si substrate (Figure 3f). The topographic AFM image in Figure 3e shows that the transferred WS$_2$ crystal is ultra-flat and follows the atomic steps of the annealed sapphire, with a step height of ~0.2 nm (Figure S2d). The observation of WS$_2$ lattice in the scanning tunneling microscope (STM) image (Figure 3f) also suggests the absence of polymer residues on the top. The field-effect transistor (FET) fabricated using the WS$_2$ layer transferred onto the SiO$_2$/Si substrate exhibits a charge carrier mobility of 1.9 cm$^2$/V·s (Figure 3g); the plateau of current saturation caused by channel pinch-off is shown in the inset. The mobility is comparable to that of high-quality CVD-grown WS$_2$ crystals at room temperature.[23, 30] These results suggest that the proposed VPS technique efficiently preserves the high quality of the WS$_2$ crystals.

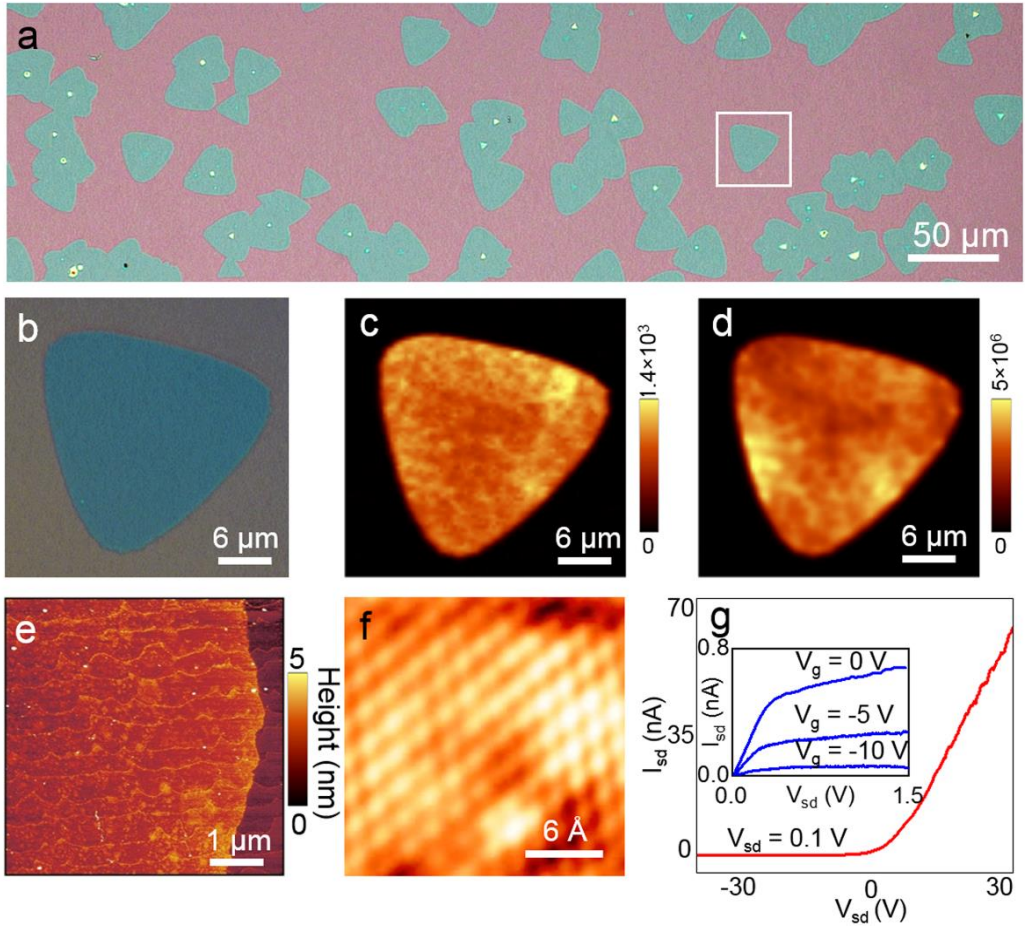

**Figure 3.** Characterization of transferred WS$_2$ single crystals. (a) Optical image of transferred WS$_2$ single crystals on SiO$_2$/Si over a large area after polymer removal. (b) Optical micrograph, (c) Raman map and (d) PL map of a representative crystal marked with white square in (a). (e) Topographic AFM image of a region of the WS$_2$ single crystal transferred on high-temperature annealed sapphire showing the ultraclean surface of the transferred crystals. (f) STM image of the transferred monolayer WS$_2$. (g) Transfer curves (drain current vs. gate voltage) of the FET fabricated using transferred WS$_2$ as the channel material on a 300-nm-thick SiO$_2$/Si substrate. The inset shows the output curves (drain current vs. drain voltage) of the fabricated device.

## 2.3 Transfer various 2D materials and fabricate heterostructures through widely used VPS technology.

The VPS technique can be used to assemble a wide range of 2D materials (other than WS$_2$) and heterostructures grown on hydrophilic surfaces. **Figure 4**a-c shows the optical image (Figure 4a), Raman map (Figure 4b), and PL map (Figure 4c) of a CVD-grown MoS$_2$ single crystal transferred from sapphire onto SiO$_2$/Si using the VPS technique. Similar to the case of WS$_2$, the uniform Raman and PL signals across the crystal indicate that the high quality of the crystal is preserved after transfer. The Raman and PL spectra of a $MoS_2$ crystal transferred on $SiO_2$/Si using this method are shown in Figure S7a and b, respectively. The VPS technique can also be used to transfer large-area in-plane heterostructures/films. Figure S8a present the optical image of the transfer of large-area stitched graphene (purple)/$WS_2$ (green) lateral heterostructures from sapphire to $SiO_2$/Si. The details for fabricating the heterostructures was reported elsewhere. [23, 30] Due to the limited quality from the commercially purchased graphene layers which contains holes and cracks prior to the growth of $WS_2$, providing nucleation sites for CVD $WS_2$ due to the existence of dangling bonds of carbon. This leads to nucleation and multilayer growth of $WS_2$ at these defects for the as-fabricated graphene/$WS_2$ lateral heterostructures, which preserves after transfer featured as lines and dots of $WS_2$ in graphene regions in Figure S8a. As shown by AFM topography (Figure S8b), the transferred heterostructures show intact graphene and $WS_2$ films and their sharp interfaces. After transfer, the $WS_2$ film retains strong PL (Figure S8c and h) and the graphene/$WS_2$ also shows uniform Raman (Figure S8d-g) signals.

VPS is useful for the direct and efficient assembly of vertical heterostructures on a large scale. For example, the vertical $MoS_2$/$WS_2$ heterojunction shown in Figure 4d and e was achieved by VPS, transferring $WS_2$ onto a sapphire substrate covered with $MoS_2$ crystals, and further transferring the heterojunctions onto a sapphire substrate. The Raman mapping and spectrum (Figure S9a and b) recorded from the $MoS_2$/$WS_2$ overlapped region shows the $E'$ peaks of $WS_2$ (350 cm$^{-1}$) and $MoS_2$ (385 cm$^{-1}$), and the $A_1'$ (405 cm$^{-1}$) peak of $MoS_2$. The PL spectrum obtained from the same point (Figure S9c) shows the direct excitonic transition energies of monolayer $MoS_2$ at the top (1.85 eV) and monolayer $WS_2$ at the bottom (2.00 eV). These values are consistent with the PL peak values obtained from individual $MoS_2$ and $WS_2$ regions (Figure S9d and e). The intensity of PL from the heterostructure region is weaker than that of monolayer $MoS_2$ or $WS_2$ regions because of the charge transfer process between $WS_2$ and $MoS_2$ layers by the formation of a type-II band alignment.[31] Owing to the clean interface between $MoS_2$ and $WS_2$, the interlayer exciton is observed at 1.42 eV[14, 32] (Figure 4f).

The VPS technique can also be used to create vdW layered structures by stacking up to four monolayers by merely repeating the stamping, peel-off, and water intercalation processes. The details of the multiple stacking are schematically shown in Figure 2a and described in Experimental Section and Figure S10 (optical images). The optical image of a four-layer $WS_2$ stack fabricated using this method are shown in Figure 4g. The low- and high-magnification topographic AFM images of the four-layer region (Figure 4h and i, respectively) shows that the surface of the stacked $WS_2$ multilayer is extremely clean, although some wrinkles are evident. We compared the VPS technique with other transfer techniques using PS or PMMA for direct stacking of multilayers using the same procedures. It was found that two layers of $WS_2$ can be stacked using only PS (Figure S10), while only one layer of $WS_2$ can be stacked using only PMMA, suggesting that PMMA is not suitable for direct staking of multilayers. These differences are probably related to the different adhesive behaviors of the polymers. The PS/PPA stamp was unable to further pick up 2D materials after the process was repeated several times. This may be attributed to the increased water from blank areas of substrate without covered 2D materials, attached to the PS/PPA stamp, which reduces the adhesion of the polymer.

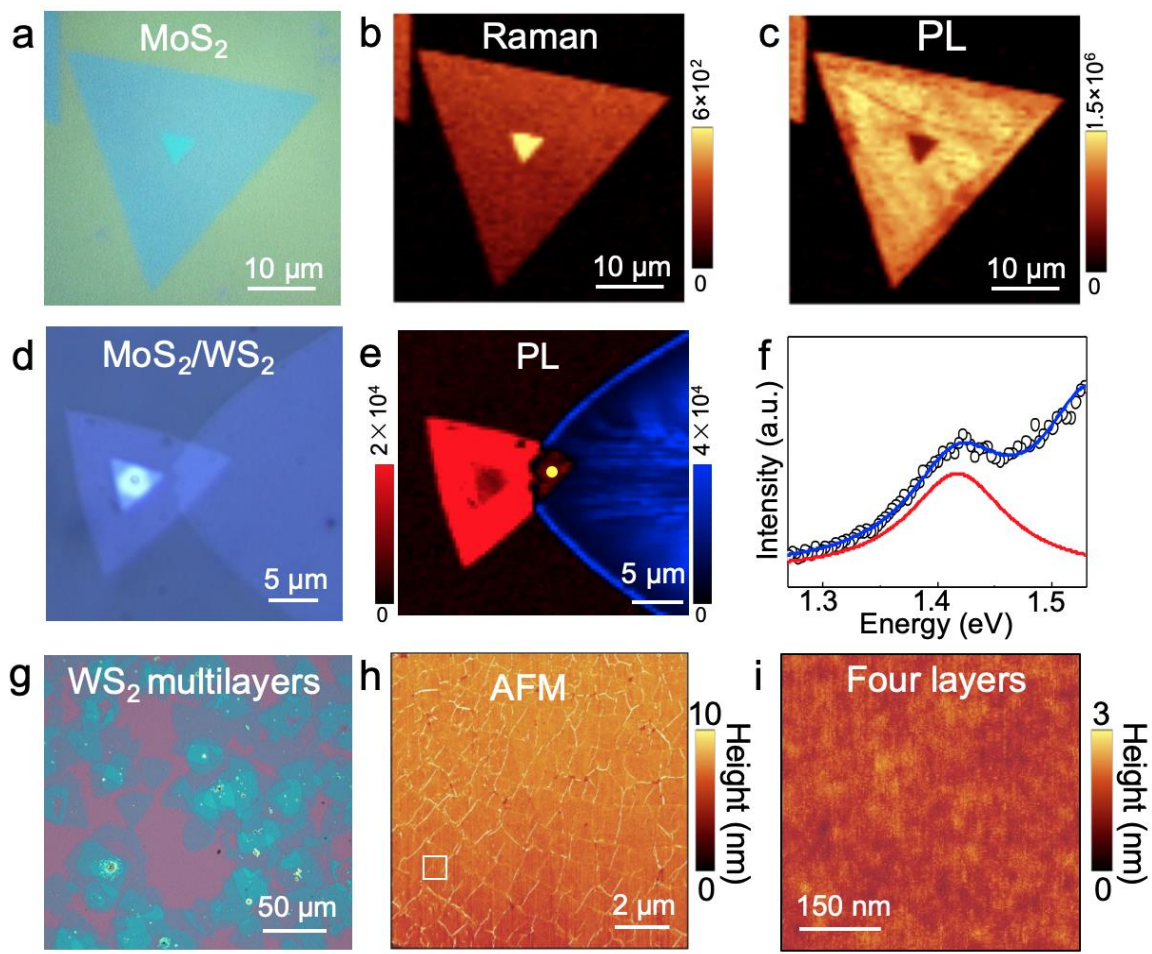

**Figure 4.** Transfer of other 2D materials and fabrication of vdW heterostructures and multilayers by VPS. (a) Optical image, (b) Raman map and (c) PL map of a CVD-grown $MoS_2$ single crystal transferred on $SiO_2$/Si. (d) Optical image of a fabricated vdW $MoS_2$/$WS_2$

heterostructure and (e) combined PL maps of the heterostructure with MoS$_2$ mapped by A1 direct excitonic transitions from 1.80 to 1.91 eV (650 to 730 nm) and WS$_2$ mapped by direct excitonic transitions from 1.91 eV to 2.10 eV (590 to 650 nm). (f) PL spectrum obtained from MoS$_2$/WS$_2$ heterostructure region (yellow dot in (e)) featuring the interlayer coupling peak at 1.42 eV. (g) Optical image of a four-layer stack of WS$_2$. (h) Low-magnification topographic AFM image of the four-layer WS$_2$ region. (i) High-magnification topographic AFM image of the region marked with white square in (h). The white features in the AFM maps in (h) are wrinkles on the multilayer WS$_2$ sheet edges induced by the multiple wet transfer process without observable features of polymer residue.

**2.4 Transfer of large-scale and suspended 2D materials by VPS technology.**

Our evaporative, non-solvent method for removing the PPA support layer enables extremely gentle transfer of 2D materials and opens up the possibility of fabricating delicate structures that are difficult to fabricate using other transfer techniques. This is because the PPA layer on top can strengthen the monolayer materials to endure large shape deformations caused by the surface tension of the liquid (such as toluene), as shown schematically in **Figure 5**a and b. In other methods, solutions such as acetone and water are widely used for polymer removal.[33-35] The monolayer materials are easily ruptured by the surface tension of the liquid during the drying process after the removal of the polymer (Figure 5c).[36]

Even though large-scale suspended 2D materials have been fabricated using complicated and delicate designs, their poor success rates and presence of polymer residues[36-38] limit their broad application. In the present method, normal solvent-washing procedures can be adopted to create freestanding 2D materials. This is because the considerably thick PS/PPA composite film exhibits improved bending stiffness. Figure 5d shows the SEM image of the monolayer WS$_2$ film on transferred on the TEM grid using our transfer method after polymer removal. Monolayer WS$_2$ is suspended over the holes (without any lacey carbon) in the TEM grid with a span larger than 150 µm (Figure 5e). To the best of our knowledge, this is the largest span reported for freestanding monolayer TMDs. The aspect ratio (length-to-thickness ratio) exceeds 200,000, which is the highest reported for suspended monolayer TMD thin films.[39, 40] This result indicates that our method is suitable for simple fabrication of freestanding 2D layers on a large scale with outstanding integrity. The PL spectrum of the suspended WS$_2$ monolayer (Figure 5f (top panel)) shows trion and exciton peaks at 1.926 and 1.963 eV, respectively via

Lorentzian fitting. Compared to the WS$_2$ crystal transferred on SiO$_2$/Si (Figure 5f, bottom panel), which shows trion and exciton peaks at 2.023 and 1.992 eV, respectively, the freestanding WS$_2$ exhibits large redshifts of 60 and 66 meV in the exciton and trion, respectively, indicating increased exciton and trion binding energies. This may be attributed to the reduced dielectric screening due to the absence of underlying substrates, which is in good accordance with theoretical calculations.[37]

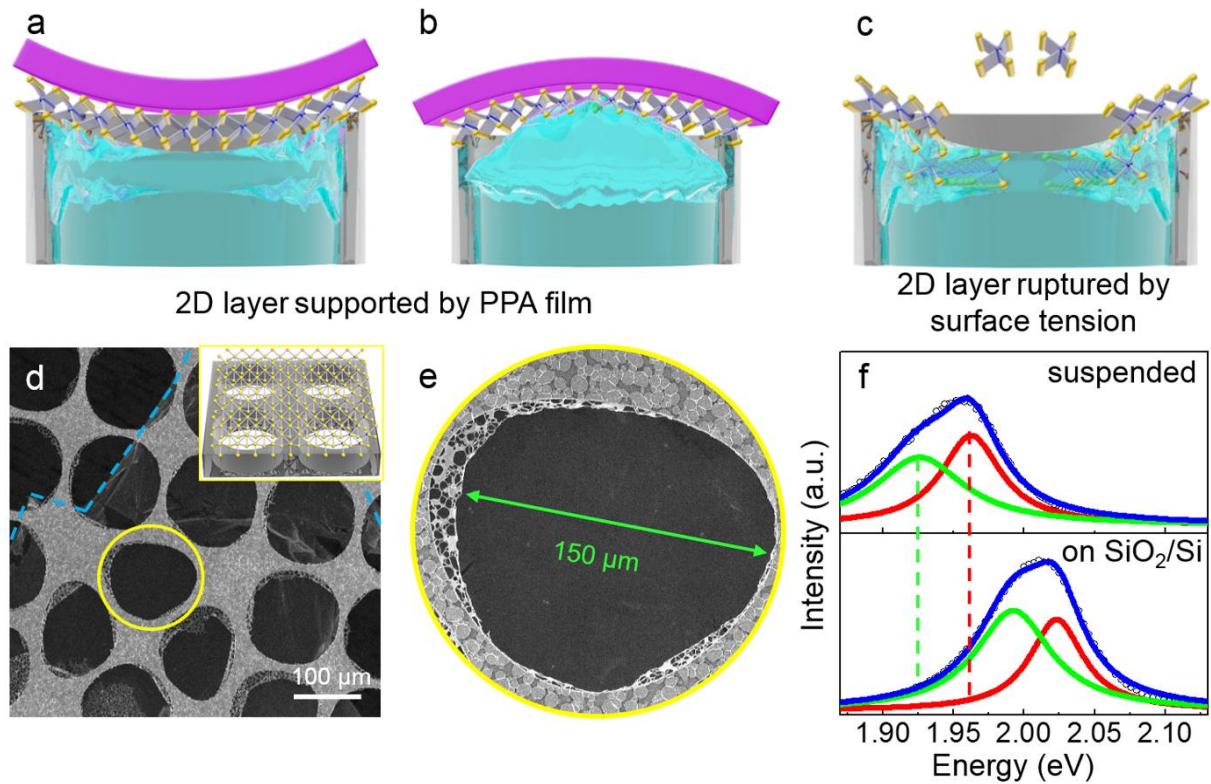

**Figure 5.** Transfer of suspended monolayer WS$_2$ film. Schematics showing (a and b) PPA film as a protective layer to prevent suspended 2D materials from rupture during the wet transfer process and (c) rupture of suspended 2D materials during wet transfer process in the absence of protective PPA film. (d) SEM image of a suspended monolayer WS$_2$ film transferred on a TEM grid. (e) High-magnification SEM image of the region marked with yellow circle in (d). (f) Comparison of PL spectra obtained from the suspended and supported WS$_2$ films.

## 3. Conclusion

We developed a versatile transfer technique using a PS/PPA composite to achieve fast, etching-free, intact, and ultraclean transfer and assembly of various 2D materials and heterostructures from hydrophilic substrates. Moreover, the polymer composite can be used as a volatile polymer stamp to assemble high-quality vdW heterolayers via simple stacking. Direct stacking enables the formation of ultraclean vdW interfaces, leading to the observation of interlayer excitons. After successful transfer and assembly, PS could be completely removed by soaking in toluene, as the solubility of PPA was negligible. The 2D materials are protected by PPA, which can be removed later by annealing at elevated temperatures. Owing to the clean and intact transfer, the FET fabricated using the transferred $WS_2$ shows mobility up to 1.9 $cm^2V^{-1}s^{-1}$, and the $MoS_2/WS_2$ vdW heterostructure exhibits interlayer coupling at 1.42 eV. The PPA layer also acts as a strong supporting layer for 2D materials during the transfer process of suspended 2D layers, enabling the fabrication of freestanding 2D materials in a gentle and efficient manner for advanced applications.

## 4. Experimental Section

*Fabrication of PS and PPA solutions*: The PS solution (90 mg/mL) was prepared by dissolving 1.8 g PS pallets ($M_w$~192,000, Sigma Aldrich) in toluene solvent (for analysis, EMSURE®) to reach a solution volume of 20 mL. The solution was allowed standing at room temperature for 24 h to ensure well dissolution of PS in toluene. The PPA solution (4%) was made by dissolving 40 mg PPA powder (cyclic, IBM) in 0.96 g anisole solvent (Sigma Aldrich). The mixture of PPA powder and anisole solvent was shaken by a vortex mixer (Lab Dancer, IKA) for 10 min. Both PS and PPA solutions were stored at 4 .

*Transfer using PS/PPA composite:*. The PPA (cyclic) in anisole solution (4%) was first spin coated on the as-grown 2D material on sapphire sample at 2000 rpm for 1 min and soft baked at 90 for 2 min. Then the sample was spin coated with PS ($M_w$~192,000) in toluene solution (90 mg/mL) at 2000 rpm for 1 min and soft baked at 80 for 3 min. Next, the edges of the spin-coated PS/PPA thin film on sapphire were scratched by a scalpel. Then the polymer composite film with 2D materials attached was delaminated by a droplet of DI water on a piece of PDMS. The detached PS/PPA/2D material composites can be picked up by a tweezer and placed on the desired target substrate (e.g. $SiO_2/Si$, TEM grid). To remove the polymers, the transferred sample was first soaked in toluene for 30 min at room temperature to wash off the PS layer. After soaking, an intact PPA layer remained on the sample, covering the transferred 2D materials. Then the sample was placed in a CVD furnace and annealed at 180 for 2 h at 0.1 Torr to volatilize the PPA layer.

*DFT calculations*: The Vienna Ab-inito Simulation Package (VASP) was used to conduct DFT calculations with the projector augmented wave method. The generalized gradient approximation (GGA) with the Perdew-Burke-Ernzerhof (PBE) functional was adopted. The kinetic energy cut-off for wavefunction expansion was set to 500eV. The effect of vdW interaction was accounted for using a dispersion-corrected PBE method. The vacuum space along Z direction was larger than 20 Å to avoid artificial interaction between periodic images. Monkhorst-Pack k-point grid was used of $7 \times 7 \times 1$ for the supercell. Atomic positions and lattice constants were fully relaxed until the difference of energy and forces were less than $10^{-4}$ eV and 0.01 eV/Å respectively. The mismatch strain between relaxed $WS_2$ monolayer and sapphire substrate is always less than 0.1%.

*Raman, PL, AFM and STM characterizations:* Raman/PL measurements were performed by a confocal microscope system (WITec alpha 300R) with a 50× objective lens and a 785 nm laser excitation for $MoS_2/WS_2$ interlayer coupling and a 532 nm laser excitation for other samples. A 600 line mm$^{-1}$ grating were used to collect the spectra. During PL and Raman mapping, 50 μW laser power and 1 s integration time were used for the prevention of sample damage. The PL intensity images were obtained via the summation of the PL intensity from 1.91 to 2.10 eV for monolayer $WS_2$ and from 1.80 to 1.91 eV for monolayer $MoS_2$. The PL peaks of excitons and trions were estimated from Lorentzian fitting. The AFM and PeakForce quantitative nanomechanical (PFQNM) measurements were carried out on a Bruker Dimension Icon in a tapping mode. The STM image was acquired by a high-temperature Arhus STM-150 (SPECS) at room temperature in ultra-high-vacuum after post-annealing treatments, with base pressure of $5 \times 10^{-11}$ mbar, Ut = 1.3 V, and It = 2.3 nA.

*FET device fabrication:* The field-effect transistor devices were fabricated on CVD grown $WS_2$ transferred on $SiO_2$/Si substrate with a bottom gate strategy. The electrodes were patterned with standard UV-light direct write lithography followed by electron-beam deposition of Ti/Au (5/50 nm) in vacuum with a chamber pressure $<6 \times 10^{-6}$ Torr. The channel length between the source and drain electrodes was ~10 μm. The drain current $I_{sd}$ as a function of bias voltage $V_{sd}$ and gate voltage $V_g$ was investigated using two-channel source meter unit (Agilent, B2902A) in ambient condition.

## Acknowledgements

We acknowledge the support from the National Natural Science Foundation of China (No. 52172162), Australian Research Council (ARC, DE140101555, DE190101249, DP150103837). Z. D. acknowledges support from the Fundamental Research Funds for the Central Universities, China University of Geosciences (Wuhan) (No. 162301202610) and the Natural Science Foundation of Guangdong Province (No. 2022A1515012145). This work was performed in part at the Melbourne Centre for Nanofabrication (MCN) in the Victorian Node of the Australian National Fabrication Facility (ANFF).

# Supplementary Material

**Table S1**. The binding energies and height of $WS_2$ on sapphire for different concentration of $H_2O$ molecules.

| $H_2O$ molecules | Concentration | Binding energy (eV) | Height (Å) |
|---|---|---|---|
| 0 | 0 | 1.145 | 2.491 |
| 1 | 0.07 | 0.981 | 3.619 |
| 7 | 0.5 | 0.965 | 4.913 |
| 14 | 1.0 | 0.949 | 5.227 |

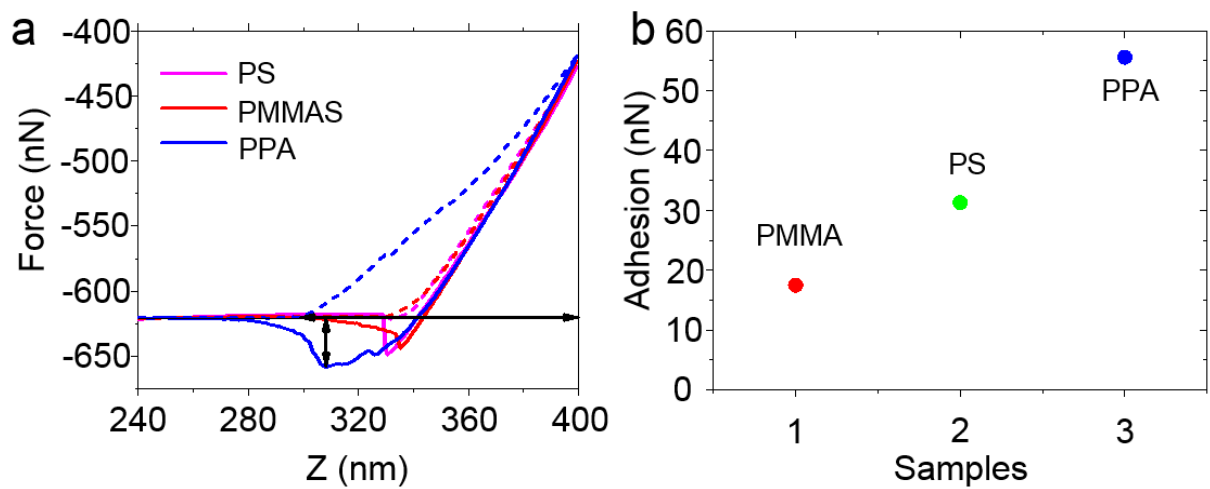

**Figure S1.** Adhesion measurement of supporting polymers (PS, PMMA and PPS) using PeakForce quantitative nanomechanical (PFQNM) technique. (a) Approach-withdraw curves of three polymers obtained by PFQNM. (b) The measured adhesion of three polymers.

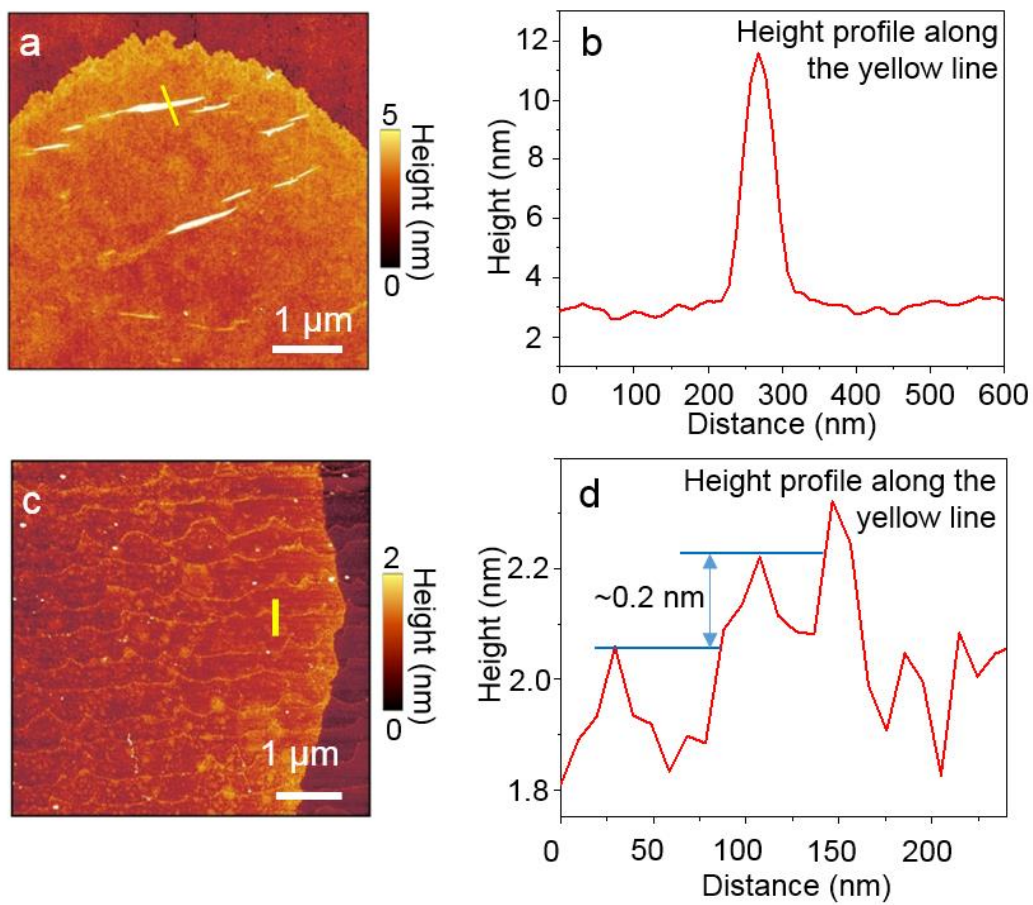

**Figure S2**. AFM height profiles of WS$_2$ single crystals transferred on SiO$_2$/Si and annealed sapphire. (a) The topographic AFM image of WS$_2$ single crystal transferred on SiO$_2$/Si from Figure 2h. (b) Height profile along the yellow line in (a). (c) The topographic AFM image of WS$_2$ single crystal transferred on high-temperature annealed sapphire from Figure 2h. (d) Height profile along the yellow line in (c).

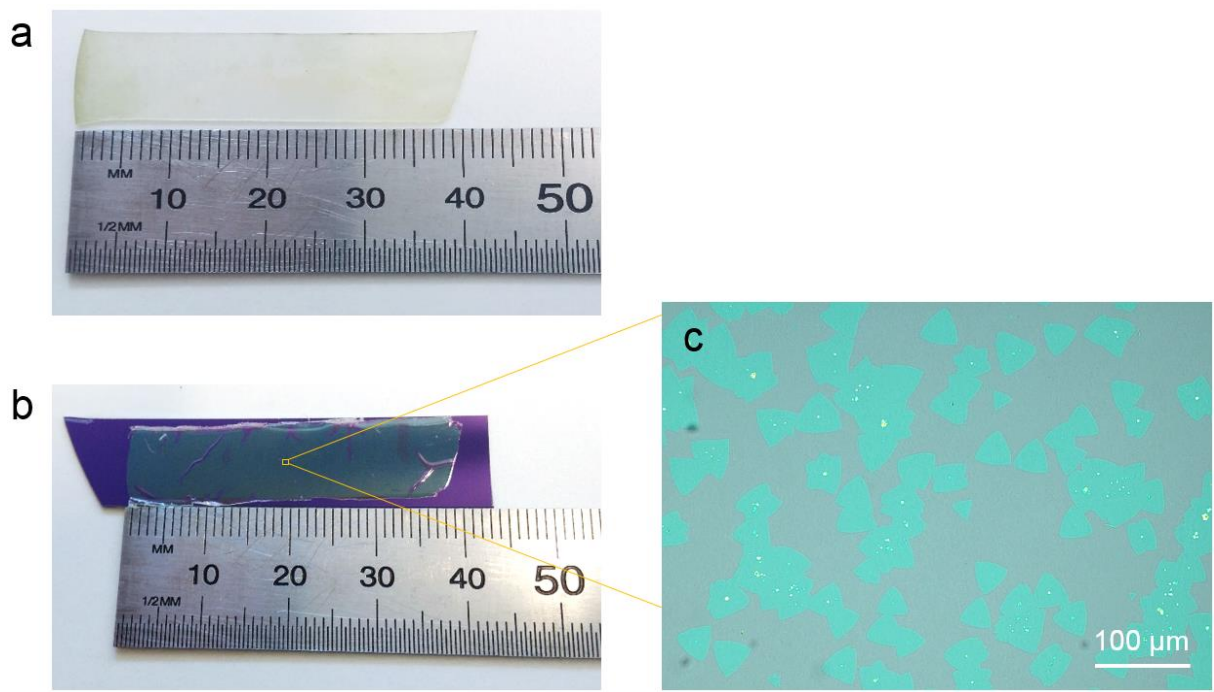

**Figure S3**. Transfer of large-scale CVD-grown monolayer WS$_2$. (a) Photo of a CVD-grown monolayer WS$_2$ on sapphire sample. (b) Photo of the transferred PS/PPA/WS$_2$ on SiO$_2$/Si. (c) Optical image of transferred WS$_2$ crystals taken from the square region in (b) after polymer removal.

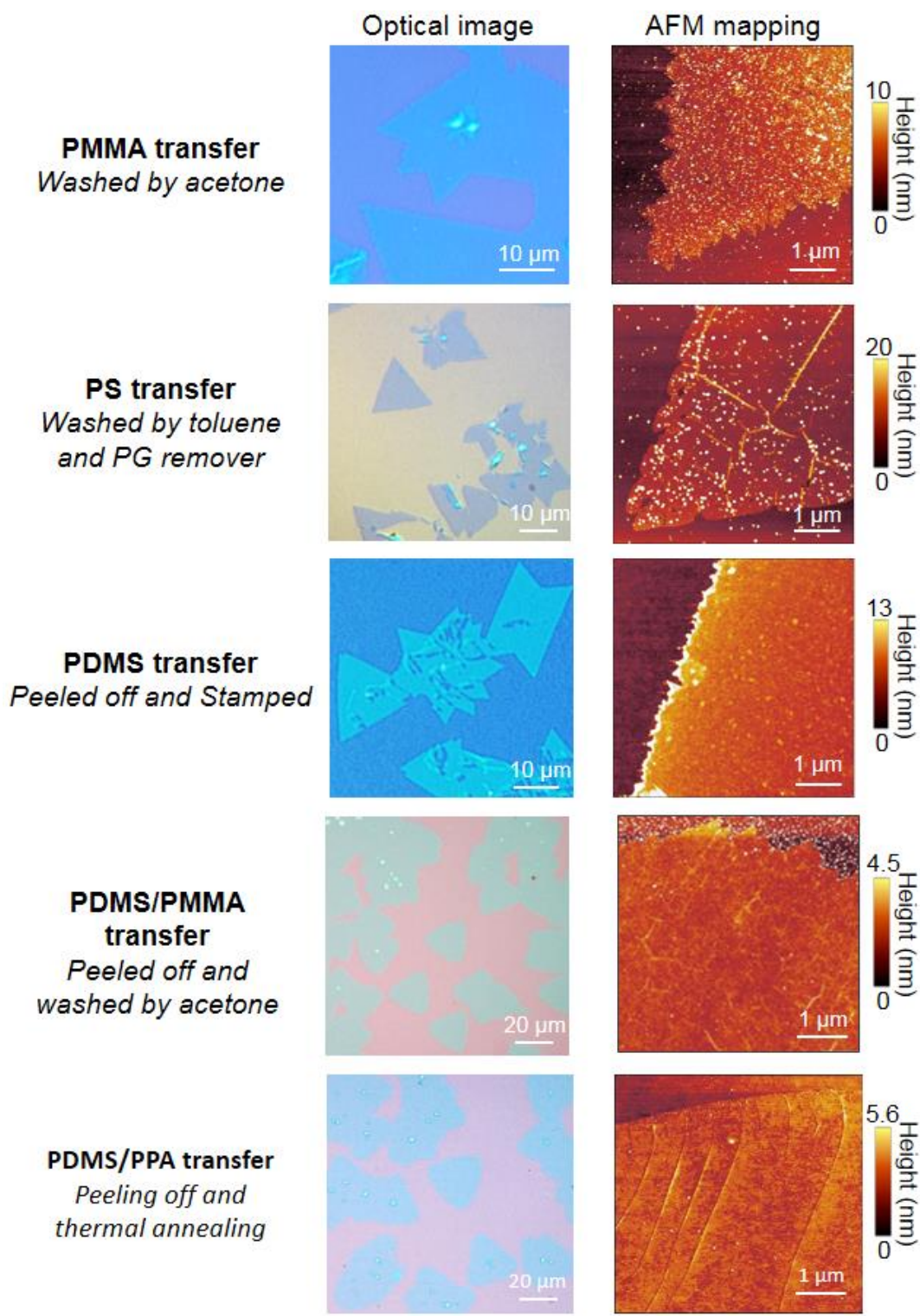

*Thermal release tap (TRT)/PMMA and TRT/PPA transfer methods do not have characterizations because only less than 10% of the polymer/2D materials can be peeled off by TRT.

**Figure S4.** Results of WS$_2$ crystals transferred on SiO$_2$/Si by other methods.

Figure S4 shows some results obtained from other transfer methods.

It can be seen from the first three cases in Figure S4 that transferring CVD grown TMDs crystals using other polymers can result in etching of the edges (*PMMA transfer*, optical and AFM images) and tearing up of the crystals (solvent stripping for *PS transfer*, optical image; peeling of the stamp for *PDMS transfer*, optical image). And all of them suffer from polymer residues as reflected by the features shown in AFM maps.

For the fourth case in Figure S4 where PMMA films are used in combination with PDMS stamps, our experimental results show that most of the crystals are not able to be delaminated from the original grown substrate, indicating extremely low efficiency of this method. This is also true for the TRT/PMMA and TRT/PPA cases. It should be noticed that there are a few crystals that can be transferred by PDMS/PMMA system in limited local regions, and these crystals were better preserved compared with the direct PDMS stamping due to the protection of PMMA (optical image). But they suffered from polymer residues upon the removal of PMMA film (AFM map).

For the last case in Figure S4 where the PDMS/PPA system is used, efficient delamination can be achieved and the surface of the crystal is relatively clean compared with other methods shown in Figure S4 due to the volatile nature of PPA. However, there are still some features observable in the AFM map indicating incomplete removal of the polymers due to residue from PDMS. And localized wrinkles are formed due to the handling and peeling of PDMS.

Also, it should be noted that none of these methods can lead to transfer of large-scale suspended 2D materials as we illustrate in Figure 5.

These results are summarized in table S2.

Table S2. Comparison of different transfer methods.

| Transfer methods | Film delamination from original substrate | | | | Polymer removal | | | Residual-free | Crystal preservation | Transfer of suspended structures |
| --- | --- | --- | --- | --- | --- | --- | --- | --- | --- | --- |
| | Alkali etching | Etch-free | | | Direct peeling-off | Solvent washing | Thermal evaporation | | | |
| | | Direct peeling-off | Water droplet intercalation | Water soaking | | | | | | |
| PMMA | ✓ | | | | | ✓ | | No | Poor[a] | Broken[d] |
| PS | | | ✓ | | | ✓ | | No | Poor[b] | Broken[d] |
| PDMS | | ✓ | | | ✓ | | | No | Poor[c] | Broken[d] |
| PDMS/PMMA | | | | ✓ | | ✓ | | No | Poor[c] | Broken[d,e] |
| PDMS/PPA | | | | ✓ | | | ✓ | No | Good | Broken[e] |
| TRT/PMMA | | ✓ | | | | ✓ | | No | Poor[c] | Broken[d,e] |
| TRT/PPA | | ✓ | | | | | ✓ | Yes | Poor[c] | Broken[e] |
| PS/PPA | | | ✓ | | | | ✓ | Yes | Good | Unbroken and good |

[a] Edges of the crystals become serrated due to alkali etching

[b] 2D materials experience tearing due to solvent stripping

[c] 2D materials experience tearing and cracking due to peeling of the stamp

[d] 2D materials are subject to breakage due to surface tension of solvents

[e] 2D materials are unable to be peeled off on large-area trench/hole structures due to non-uniform contact surfaces

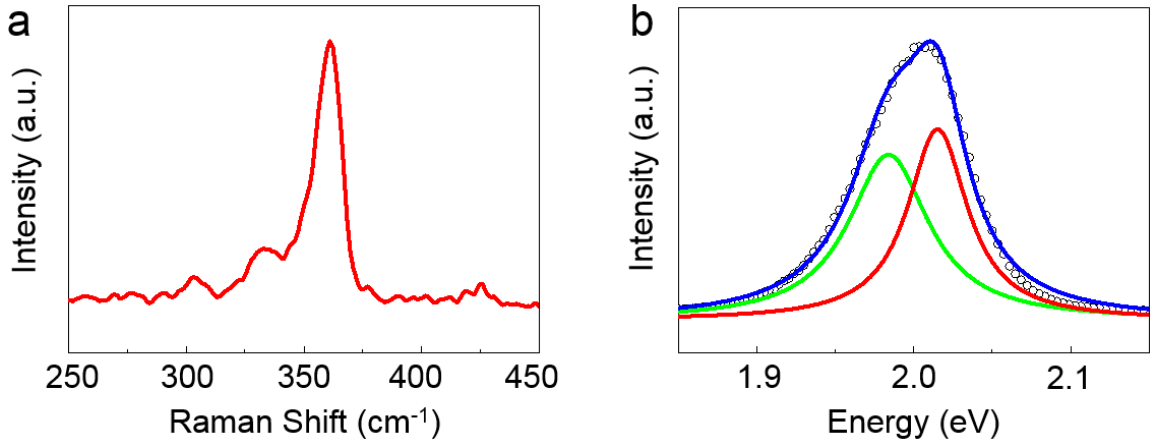

**Figure S5.** (a) Raman and (b) PL spectra of the $WS_2$ crystal transferred on $SiO_2$/Si using VPS technique corresponding to Figure 3c and d. The characteristic Raman E peak at 350 cm$^{-1}$ for monolayer $WS_2$ can be observed in the Raman spectrum in (a). The Lorentz fitting of the PL peak for the as-transferred $WS_2$ indicates an exciton peak located at 2.015 eV and a trion peak located at 1.984 eV as shown respectively by the green and red curves in (b), comparable with the values reported by Ref.[1] where the CVD grown $WS_2$ single crystals were transferred on $SiO_2$/Si.

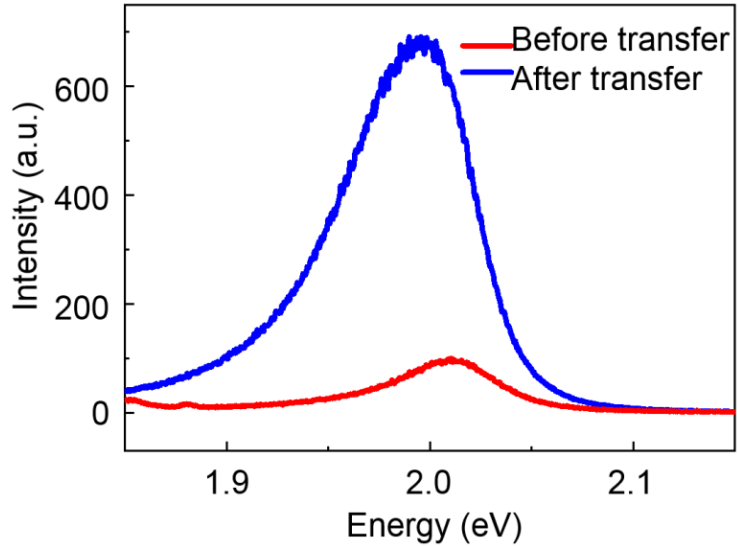

**Figure S6.** PL spectra of $WS_2$ crystal as-grown on sapphire (red) and transferred on another piece of sapphire (blue). Same laser power was used for obtaining the spectra in both cases. The PL intensity value is normalized for $WS_2$ before transfer and was scaled accordingly for the $WS_2$ after transfer.

It can be seen that the intensity of PL for the $WS_2$ after transfer is ~7 times stronger compared with when as-grown on sapphire. As same type of substrate is used, this strong PL of transferred $WS_2$ indicated excellent $WS_2$ crystal quality after transfer. The increase in PL intensity after transfer can be attributed to the influence of trapped water beneath the transferred $WS_2$ which cannot be effectively removed from the hydrophilic surface of the transferred sapphire substrate by the annealing process at 180 °C. This trapped layer of water molecules serves as a good dielectric which screens the electron accepting nature of sapphire surface and recovers the intrinsic n-type doping of CVD grown $WS_2$.[2] This also accounts for the redshift of the PL peak after transfer. The negative impact of the trapped water on the electrical performance of the 2D materials can be removed by annealing at high source-drain bias [3] or prolonged baking in air [4].

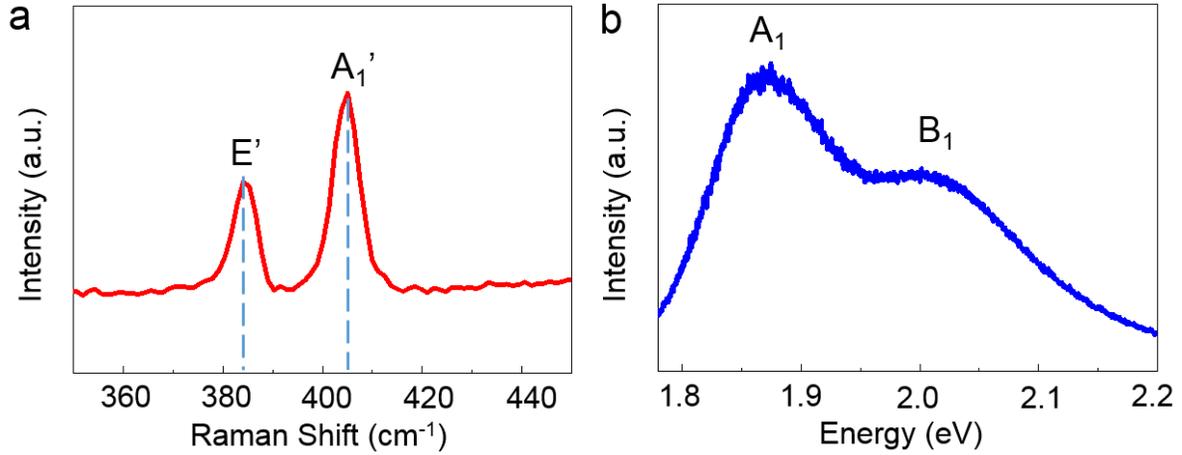

**Figure S7.** (a) Raman and (b) PL spectra of a MoS$_2$ single crystal transferred on SiO$_2$/Si. The Raman spectrum features the characteristic in-plane *E'* peak at ~385 cm$^{-1}$ and out-of-plane *A$_1$'* peak at ~ 405 cm$^{-1}$ for monolayer MoS$_2$.[5] While the PL spectrum exhibits A1 and B1 exciton peaks at 1.866 and 2.011eV, respectively, blueshifted compared to Ref. [5] and [6] where MoS$_2$ was directly grown on SiO$_2$/Si due to strain release after transfer.

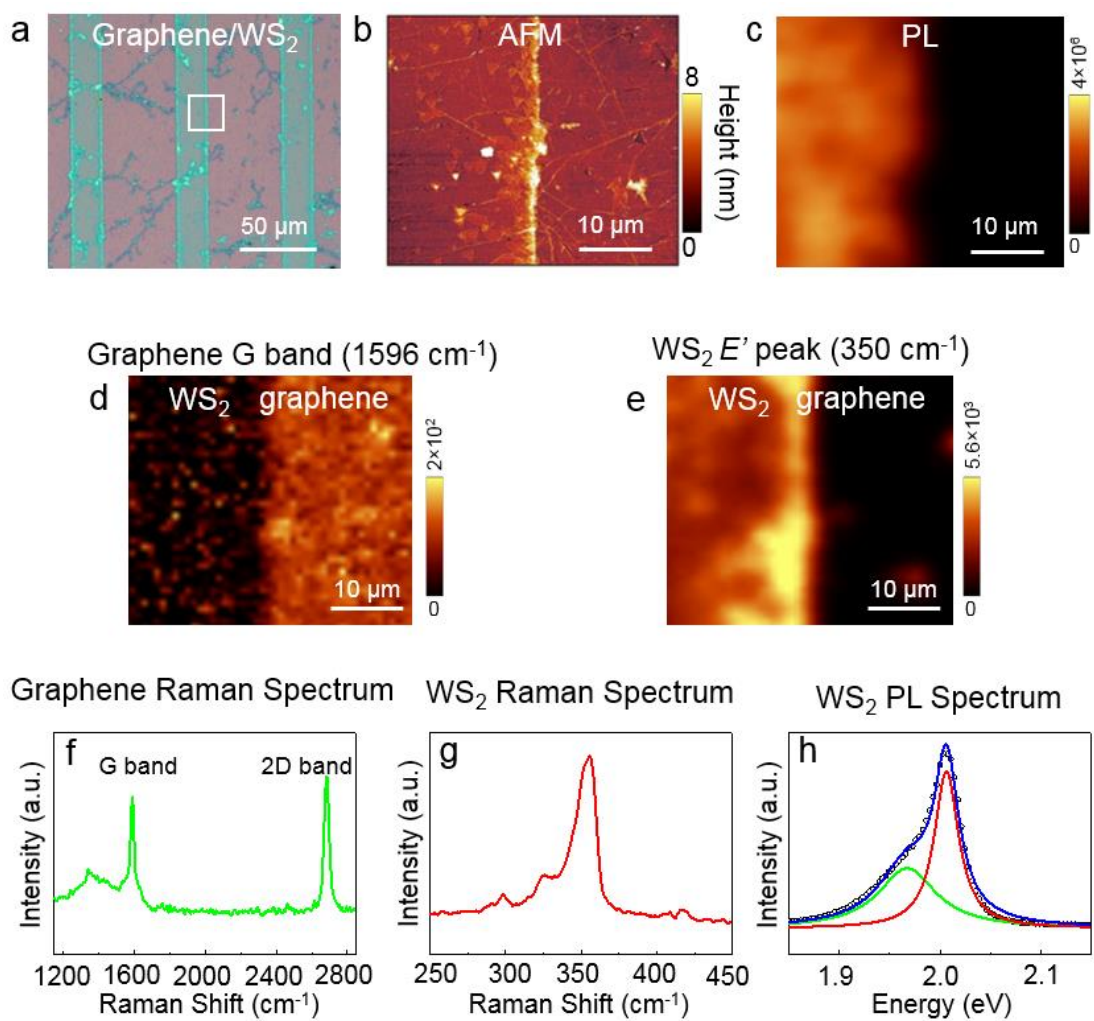

**Figure S8**. Transfer of graphene/WS$_2$ in-plane heterostructures by VPS and AFM and optical characterizations. (a) Optical image of large-area in-plane graphene/WS$_2$ heterostructures transferred on SiO$_2$/Si. (b) AFM topograph and (c) PL map of the square region in (a). The white features in (b) are multilayer WS$_2$ growth in the heterostructure. (d) graphene G band and (e) WS$_2$ *E'* peak of graphene/WS$_2$ in-plane heterostructure transferred on SiO$_2$/Si. (f) Raman spectrum of graphene collected in Figure S9a. The sharp G band (1596 cm$^{-1}$) and 2D band (2687 cm$^{-1}$) indicate the monolayer nature of the graphene. There is also a broad peak observed at the position of D band (~1350 cm$^{-1}$) which has been identified before the transfer process in our previous work on the graphene/WS$_2$ heterostructures and has been attributed to the doping effect from the PMMA residue.[7] This may also cause the 2D peak to G peak intensity ratio being less than 2.[8] (g) Raman spectrum of WS$_2$ collected in Figure S9b indicating characteristic *E'* Raman peak of monolayer

WS$_2$ at 350 cm$^{-1}$. (3) PL spectrum of WS$_2$ collected in (c). The Lorentz fitting of the PL peak shows the trion peak (green) at 1.967 eV and the exciton peak (red) at 2.006 eV. The PL peaks are much less blueshifted compared with results shown in Figure S5 due to the stitching effect of the in-plane heterostructures leading to less strain released after transfer.

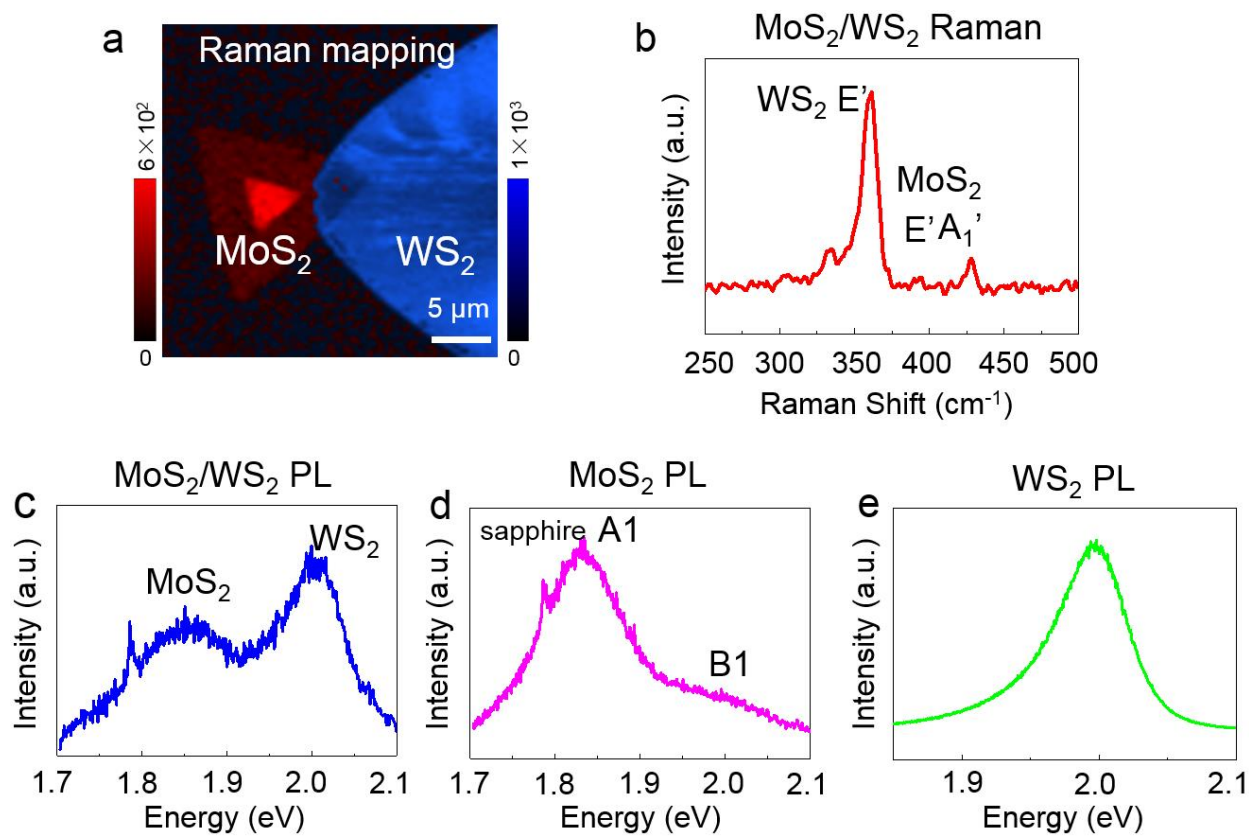

**Figure S9.** (a) Raman mapping of the MoS$_2$/WS$_2$ van der Waals heterostructure in Figure 4d excited by 532 nm laser. The red region corresponds to MoS$_2$ Raman A$_1$' peak mapped at 410 cm$^{-1}$ and the blue region corresponds to WS$_2$ Raman E' peak mapped at 360 cm$^{-1}$. (b) Raman and (c) PL spetra on the overlapping region of the MoS$_2$/WS$_2$. (d-e) PL spectra of (d) MoS$_2$ and (e) WS$_2$ taken from their individual regions in Figure 4e. The PL peaks values observed separately for MoS$_2$ and WS$_2$ at their individual regions are consistent with the PL peaks identified at the MoS$_2$/WS$_2$ heterostructure region.

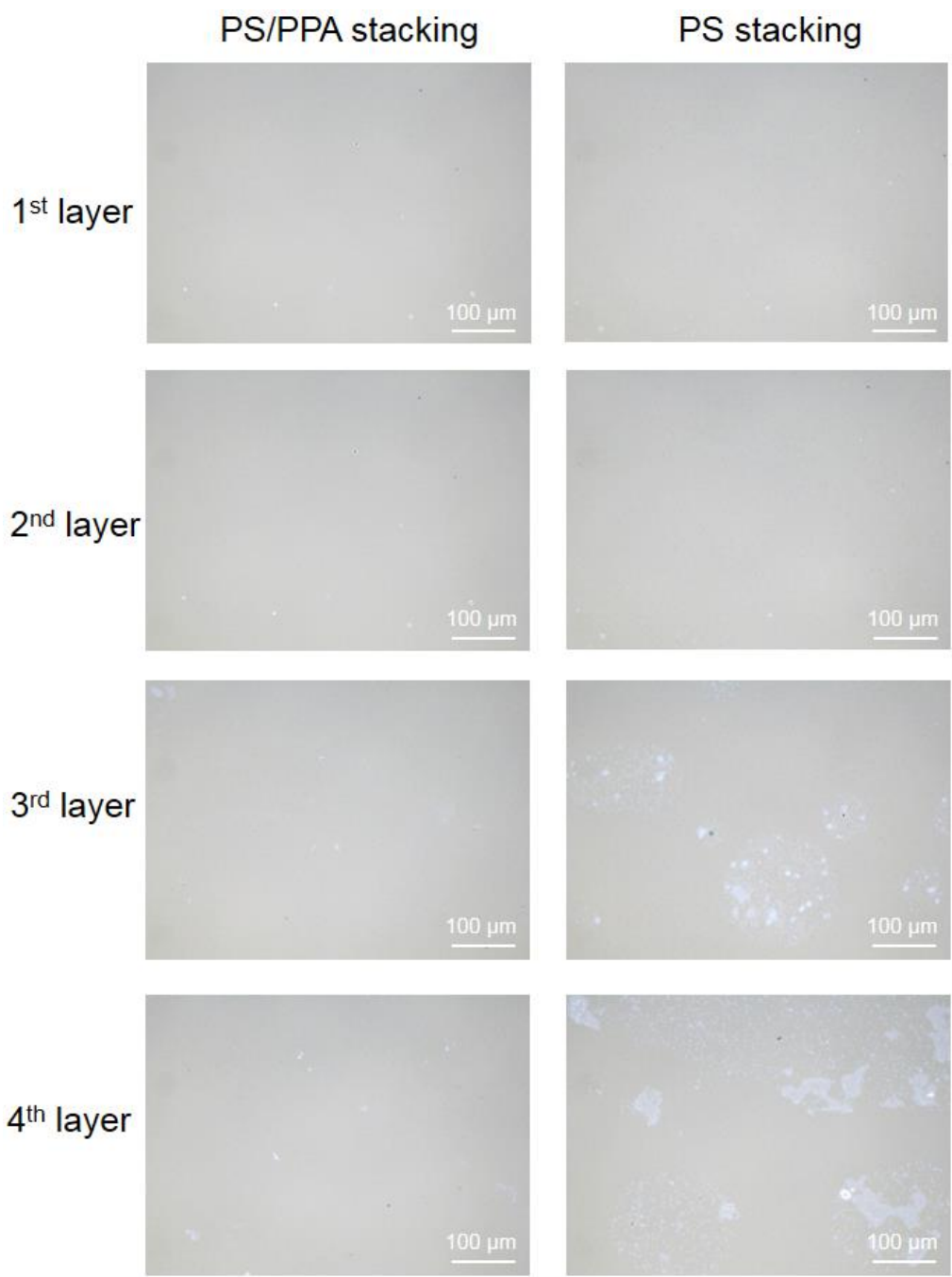

**Figure S10.** Optical images of sapphire surface after the as-grown WS$_2$ being picked up by stamps of PS/PPA in comparison with PS, indicating different picking-up efficiencies.

We use *PS/PPA/n layers of WS$_2$* (n = 0, 1,2, …) as a stamp to directly stack multilayer WS$_2$, as demonstrated in Figure 1a. We then image the surface of the substrate after the originally grown

WS$_2$ has been picked up by this method using an optical microscope. Thus, the cleanness of the imaged surface reflects the pickup efficiency of the stamp. To be more specific, we pick up the first layer of WS$_2$ crystals that are CVD grown on sapphire (Sample #1) using the *PS/PPA* stamp. After this pickup process, the as-grown WS$_2$ is attached to the PS/PPA polymer composites and the surface of the sapphire substrate of Sample #1 after the pickup process is imaged and shown in "PS/PPA stacking, 1$^{st}$ layer" panel of Figure S9. There is no WS$_2$ crystals left on the substrate, indicating that all the crystals have been picked up by PS/PPA. Then we adopt the obtained PS/PPA attaching the 1$^{st}$ layer WS$_2$ as the stamp and perform this pickup process on another piece of WS$_2$ CVD grown on sapphire sample (Sample #2). Similarly, the WS$_2$ crystals from this second sample are picked up by the *PS/PPA/1 layer of WS$_2$* stamp and the surface of the sapphire substrate of Sample #2 after the pickup process is imaged and shown in "PS/PPA stacking, 2$^{st}$ layer" panel of Figure S9. The surface of the sapphire for this second sample after pickup is also clean with no WS$_2$ remaining, indicating high pickup efficiency with PS/PPA stamp attaching an additional WS$_2$ layer. We do this continuously on four WS$_2$ on sapphire samples using both PS/PPA and PS along and obtain the Figure S10.

It can be seen that for picking-up using PS/PPA stamp, the sapphire substrates are still clean with few remaining WS$_2$ after the picking-up of the fourth WS$_2$ on sapphire sample, indicating high picking-up efficiency by PS/PPA. In contrast, for picking-up with PS only, there are much more WS$_2$ remaining on the sapphire substrates after the picking-up of the third and fourth WS$_2$ on sapphire samples, indicating low picking-up efficiency by PS.

**Supporting References**